# Heterogeneous nucleation of/on nanoparticles: a density functional study using the phase-field crystal model

László Gránásy,[*,a,b] Frigyes Podmaniczky,[a] Gyula I. Tóth,[a] György Tegze[a] and Tamás Pusztai[a]

Crystallization of supersaturated liquids usually starts by heterogeneous nucleation. Mounting evidence shows that even homogeneous nucleation in simple liquids takes place in two steps; first a dense amorphous precursor forms, and the crystalline phase appears via heterogeneous nucleation in/on the precursor cluster. Herein, we review recent results by a simple dynamical density functional theory, the phase-field crystal model, for (precursor-mediated) homogeneous and heterogeneous nucleation of nanocrystals. It will be shown that the mismatch between the lattice constants of the nucleating crystal and the substrate plays a decisive role in determining the contact angle and nucleation barrier, which were found to be non-monotonic functions of the lattice mismatch. Time dependent studies are essential as investigations based on equilibrium properties often cannot identify the preferred nucleation pathways. Modeling of these phenomena is essential for designing materials on the basis of controlled nucleation and/or nano-patterning.

[a] *Wigner Research Centre for Physics, P.O.Box 49, H−1525 Budapest, Hungary*
[b] *BCAST, Brunel University, Uxbridge, Middlesex, UB8 3PH, UK*
[*] *Corresponding author: granasy.laszlo@wigner.mta.hu*

**Key learning points**
(1) Homogeneous crystal nucleation often takes place via metastable (amorphous or crystalline) precursors.
(2) The mismatch of the lattice constants of the nucleating crystal and the substrate is an essential parameter: the contact angle, thickness of adsorbed crystal layer, and the nucleation barrier are non-monotonic functions of the lattice mismatch.
(3) The free growth limited model of particle induced crystallization by Greer and coworkers is successful for small anisotropies and supersaturations, whereas it fails for faceted crystals or small nanoparticle agents.
(4) Large lattice mismatch may lead to the formation of an amorphous surface layer.
(5) Time dependent studies are essential for finding the nontrivial nucleation pathways.

## 1. Introduction

When a liquid is cooled below its melting point, it is no longer stable and freezes eventually. The liquid exists in the metastable undercooled state until a nucleation event occurs, during which the new phase appears via crystallike fluctuations termed "heterophase" fluctuations. If the heterophase fluctuations exceed a critical size (usually determined by the driving force of crystallization, the solid-liquid interface free energy, its anisotropy, and possible contact to heterogeneities), they grow further with a high probability, whereas the smaller ones tend to decay. Heterophase fluctuations of the critical size are termed the *critical fluctuations or nuclei*, and the respective work of formation is the *thermodynamic barrier of nucleation*, which the system needs to pass via fluctuations to reach the bulk crystalline state. The nucleation process may be either *homogeneous* or *heterogeneous*. Homogeneous nucleation takes place in an idealized supersaturated liquid, where the internal fluctuations of the liquid lead to the passing of the thermodynamic barrier of formation of crystallites. In turn, heterogeneous nucleation occurs in "impure" liquids, in which heterogeneities, such as container walls or nucleating agents (termed here "substrate") are introduced to the melt (either intentionally or not), which facilitate nucleation via reducing the free energy barrier to the formation of the crystal.[1(a)] This reduction happens, when the substrate induce ordering in the liquid that helps the formation of the crystalline phase. Heterogeneous nucleation is not only a phenomenon of classic importance in materials science but attracts continuously growing interest due to the emerging technological interest in micro- and nanopatterning techniques,[2] and the control of related nanoscale processes, such as crystallization on patterned substrates, including the formation of quantum dots,[3] the properties of glass ceramics produced by controlled nucleation,[1(b)] phase selection in alloys,[4] copper nucleation on graphene,[5] and the undercoolability of living organisms,[1(c)] to mention a few examples. Despite its technological importance, heterogeneous nucleation is relatively little understood owing to difficulties in describing the interaction between the foreign matter and the solidifying melt.

In classical theory of heterogeneous nucleation,[1(a)] the effect of the heterogeneity in enhancing or suppressing the solid phase is formulated in the language of wetting. Having the interface free energies of the liquid-solid ($\gamma_{SL}$), wall-liquid ($\gamma_{WL}$), and wall-solid ($\gamma_{WS}$) boundaries, one may calculate the contact angle at a solid-liquid-wall triple junction (assuming isotropic interface free energies) using the Young-Laplace equation,

$$\cos(\theta) = \frac{\gamma_{WL} - \gamma_{WS}}{\gamma_{SL}}. \qquad (1)$$

In this framework, the surface is wet by the solid phase for $\theta = 0$, i.e., there will be no barrier to crystal nucleation, whereas for $\theta = \pi$ the liquid phase is preferred at the interface. According to the classical "spherical cap" model, the nucleation barrier is simply reduced by the catalytic potency factor $f(\theta)$: $W_{hetero} = W_{homo} f(\theta)$, where $f(\theta) = [\theta - \frac{1}{2} \sin(2\theta)]$ and $f(\theta) = \frac{1}{4}[2 - 3\cos(\theta) + \cos(\theta)^3]$ for 2D and 3D, respectively; i.e., only that part of the (circular / spherical) homogeneous nucleus needs to be formed by thermal fluctuations, which realizes the appropriate *contact angle* at the perimeter (see Fig. 1). The contact angle is an input for a variety of field theoretic models of the liquid-solid-wall trijunction.[6]

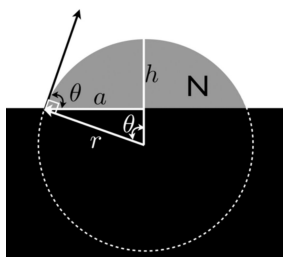

**Fig. 1** Classical "spherical cap" approach to heterogeneous nucleation on a flat surface. The white dotted line shows the contour of the homogeneous nucleus, the grey area is the heterogeneous nucleus (N). (White – liquid; black – substrate; and grey – crystal; $a$ – radius of contact surface; $h$ – height of nucleus; $r$ – radius of homogeneous nucleus; and $\theta$ – contact angle.)

The efficiency of the heterogeneities in reducing the thermodynamic barrier of nucleation is influenced by a range of microscopic properties including the crystal structure, lattice mismatch, surface roughness, surface precipitates, adsorption, etc., all requiring atomistic description. Recent molecular dynamics and Monte Carlo simulations have addressed the interaction between a foreign wall and crystallizing fluid.[7-9] The (111) face of the hard-sphere crystal wets the unstructured wall nearly ideally, and the results can only be interpreted if a line tension is also considered,[7] a finding reproduced by the lattice gas model.[8] Crystallization on substrates of triangular and square lattices, or of zig-zag stripe and rhombic patterns have been investigated.[9] Of these, the first three patterns can be matched by cutting the fcc crystal along the (111), (100), and (110) planes, whereas the rhombic pattern corresponds to a sheared fcc crystal. The simulations indicate that even for perfectly matching lattice constants, complete wetting occurs for only (111) pattern, but not for (110) and (100). In these studies, the crystallization happens via a layer-by-layer, where the first crystalline layer forms well below the bulk crystallization transition. For the rhombic pattern either incomplete wetting by only a few layers takes place, or there is no wetting at all. In the case of the triangular pattern, crystallization of the first layer is promoted if the lattice constant of the substrate is larger than that of the coexisting bulk crystal.[9] The presence of the substrate/wall leads to structural ordering in the adjacent liquid layers,[10] a phenomenon that may influence the adsorption of crystalline molecule layers at the surface of the substrate.

These findings are of primary importance from the viewpoint of a recent highly successful *free-growth limited model of particle-induced crystallization* proposed by Greer and coworkers;[1,11] a model in which cylindrical particles, whose circular faces (of radius $R$) are ideally wet by the crystal, remain dormant during cooling until the radius of the homogeneous nuclei becomes smaller than $R$, and free growth sets in. [The critical undercooling is $\Delta T_c \approx 2\gamma_{SL}/(\Delta s_f R)$, where $\Delta s_f$ is the volumetric entropy of fusion.] This mechanism has already been addressed in a coarse-grained phase-field model (see Fig. 2),[6(a)] however, atomic scale modeling would be important to understand the limitations of this essential model of initiating crystallization.

Finally it is worth mentioning that mounting evidence indicates that homogeneous nucleation is often a two-stage process, in which the stable phase appears via a metastable precursor (an intrinsic heterogeneity), a process that can be regarded as a specific heterogeneous nucleation process from the viewpoint of the stable phase. An early analysis of Alexander and McTague suggested that crystallization to the bcc phase is preferred in simple liquids.[12] A recent reiteration of the problem in terms of density functional theory concludes that the bcc phase should rather be the phase that nucleates.[13] This prediction is consistent with molecular dynamics simulations for the Lennard-Jones system (where the stable phase is fcc), where the subcritical crystalline fluctuations have the metastable bcc structure, while the critical fluctuation has an fcc core surrounded by a bcc-like surface layer.[14]

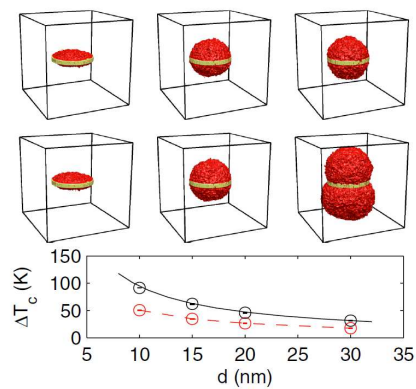

**Fig. 2** Conventional (coarse-grained) phase-field simulations illustrating the *free growth limited mode of particle induced crystallization* of pure Ni.[9(a)] Cylindrical particles ($d$ = 20 nm) with contact angles of 45° and 175° on the horizontal and vertical surfaces were used. Upper row: $\Delta T$ = 26 K < $\Delta T_c$. Central row: $\Delta T$ = 27 K > $\Delta T_c$. Time elapses from left to right. Bottom row: $\Delta T_c$ vs particle diameter $d$. Original theory — solid line; phase-field simulation — dashed line. The deviation in $\Delta T_c$ between theory and simulations is due to the thermal fluctuations considered in the latter. (Reprinted with permission from L. Gránásy, T. Pusztai, D. Saylor and J. A. Warren, *Phys. Rev. Lett.*, 2007, **98**, art. no. 035703 © 2007 American Physical Society.)

Composite bcc/fcc nuclei have also been predicted by the density functional theory[15] and a Ginzburg-Landau free energy based phase-field theory.[4] Experiments on globular proteins have shown that a metastable critical point in the supersaturated liquid may help the formation of crystal nuclei via liquid phase separation, leading to composite nuclei of crystal surrounded by dense liquid,[16] a finding recovered by computer simulations[17] and density functional/phase-field computations.[18] Brownian dynamics simulations indicate the formation of medium range crystalline order in the supersaturated liquid preceding crystallization.[19] Recent experiments on colloidal systems (Fig. 3),[20] and theoretical studies and computer simulations for simple liquids (Lennard-Jones[21] or hard-sphere[22]) imply that the appearance of the crystalline phase is preceded by an amorphous/dense liquid precursor. These results imply that precursor assisted crystal nucleation is a fairly general phenomenon creating nanocrystals in an essentially heterogeneous manner.

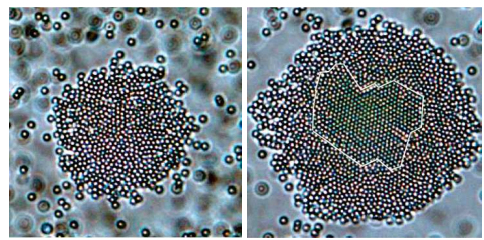

**Fig. 3** Amorphous precursor mediated crystal nucleation in 2D polymeric system (polystyrene spheres of diameter 0.99 μm and polydispersity <5%, in deionized water). (Reprinted with permission from T. H. Zhang and X. Y. Liu, *J. Am. Chem. Soc.*, 2007, **129**, 13520–13526 © 2007 American Chemical Society.)

Systematic studies of homogeneous[23] and heterogeneous[24] crystal nucleation have recently been performed using a simple dynamical density functional theory, termed the Phase-Field Crystal (PFC) model,[25,26] representing a system of Deryaguin-Landau-Verwey-Overbeek-type (DLVO) interaction,[23] in which several crystalline phases (bcc, hcp, and fcc) compete with the amorphous phase during crystallization.[27] As opposed with other atomistic approaches, the PFC model works on a diffusive time scale and can be regarded complementary to molecular dynamics.[27] Time dependent PFC simulations have shown that in the case of homogeneous nucleation the density and structural changes decouple beyond a critical undercooling/supersaturation, leading to amorphous precursor mediated crystallization;[23] the effect of lattice mismatch on the heterogeneous process[24] has also been investigated.

In this Tutorial Review, we present recent advances PFC modeling of heterogeneous nucleation of nanocrystals has made. The PFC results partly support the results obtained by other atomistic methods; partly they are complementary. The structure of our review is as follows: In Section 2, we recall briefly the main features of the PFC models, including the predicted phase diagram and homogeneous nucleation (including the two-step mechanism of homogeneous nucleation via an amorphous precursor). Section 3 addresses heterogeneous nucleation and crystal adsorption on flat walls and on crystalline particles of various shapes (cube and pitted wall) and structures (simple cubic and fcc), while varying the lattice mismatch between the nucleating crystal and the substrate. In Section 4, we give a summary of the results and offer a few concluding remarks.

## 2. PFC models for crystal nucleation

The phase-field crystal (PFC) models can be considered as simple classical dynamic density functional theories (DDFT). The local state of matter is characterized by a time averaged number (or particle) density field that depends on time and position. This time averaged number density is homogeneous in the liquid, whereas density peaks appear in the crystal at the atomic sites. Variants of the PFC model differ in the form of the free energy functional and the equation of motion.[26] The equilibrium properties, such as the interface free energy and the phase diagram can be evaluated using the Euler-Lagrange equation. In this Section we briefly recapitulate the essence of the PFC models used in nucleation studies. Since a recent review[26] covers most of the important details of the PFC models; herein, we review only the minimum information needed to understand the results presented. For further details regarding PFC modeling, the interested reader should see Ref. 26.

### 2.1 Free energy functionals

*(a) Single-mode PFC model:* The earliest version of the PFC model has been developed by Elder *et al.*[25] It is also known as the single-mode phase-field crystal model, and relies on the following free energy functional

$$\Delta F = \int d\mathbf{r} \left\{ \frac{\psi}{2} \left[ -\varepsilon + (1+\nabla^2)^2 \right] \psi + \frac{\psi^4}{4} \right\}, \quad (2)$$

where $\Delta F$ is the dimensionless (Helmholtz) free energy difference counted relative to a homogeneous reference liquid, which transforms into the dimensional free energy as follows: $\Delta \mathcal{F} = (3\rho_L^{ref} kTR^d B_s^2) \cdot \Delta F$. Here $\rho_L^{ref}$ is the particle density of the reference liquid, $k$ Boltzmann's constant, $T$ the temperature, $R$ the length scale [corresponding to the position of the peak of the direct correlation function $C(q)$], and $B_s = K/(\rho_L^{ref} kT)$, whereas $K$ is the bulk modulus of the crystal. $\psi$ is the reduced particle density, $\mathbf{r}$ the dimensionless position vector, while $\varepsilon < 0$ is the distance from the critical point in the system (located at $\psi = 0$, $\varepsilon = 0$). Parameter $\varepsilon$ is the reduced temperature, which can be related to the compressibility of the liquid, the bulk modulus of the solid, and the lattice constant. This form of the free energy can be derived from the perturbative density functional approach of Ramakrishnan-Yussouff,[28] via simplifications that include the expansion of the two-particle direct correlation function in Fourier space up to $4^{th}$ order.[29] The approximations lead to a *well defined wavelength* for the particle density, which is preferred by the system (hence the name 'single-mode' PFC). As a result, any periodic density distribution that is consistent with this wavelength represents a local minimum of the free energy. Accordingly, elasticity and crystal anisotropies are automatically included into the model. The phase diagrams, the single-mode PFC model has in 2D and 3D, are shown in Figs. 3(a) and 3(b). The PFC model has been used successfully to address a broad range of phenomena[26] including elasticity and grain boundary dynamics, the anisotropy of the interfacial free energy and the growth rate, dendritic and eutectic growth, glass formation, melting at dislocations and grain boundaries, polymorphism, and colloidal crystal aggregation.

*(b) Two-mode PFC model:* An attempt has been made to formulate a free energy functional that prefers the fcc structure at small $\varepsilon$,[30] where a linear elastic behavior persists. To realize this, two wavelengths were used (first and second neighbor reciprocal lattice vectors), hence the name "two-mode PFC" model. The corresponding free energy functional reads as

$$\Delta F = \int d\mathbf{r} \left\{ \frac{\psi}{2} \left[ -\varepsilon + (1+\nabla^2)^2 \{R_1 + (Q_1^2 + \nabla^2)^2\} \right] \psi + \frac{\psi^4}{4} \right\}. \quad (3)$$

There are two new model parameters: $R_1$ controls the relative stability of the fcc and bcc structures, whereas $Q_1$ is the ratio of the two wave numbers [for fcc, $Q_1 = 2/\sqrt{3}$ using the reciprocal lattice vectors (111) and (200)]. Note that the single-mode PFC can be recovered in the limit $R_1 \to \infty$. The phase diagrams for $R_1 = 0$ and $R_1 = 0.05$ are shown in Figs. 4(a) and 4(b).

The free energies of the single- and two-mode PFC models can be given in a unified form, which interpolates between the two limiting cases by varying parameter $\lambda = R_1/(1+R_1)$ from 0 to 1:

$$\Delta F = \int d\mathbf{r} \left\{ \frac{\psi}{2} \left[ -\varepsilon + (1+\nabla^2)^2 \{\lambda + (1-\lambda)(Q_1^2 + \nabla^2)^2\} \right] \psi + \frac{\psi^4}{4} \right\}. \quad (4)$$

Here, $\lambda = 0$ recovers the two-mode PFC model ($R_1 = 0$), whereas the limit $\lambda = 1$ yields the single-mode model.

In both models a spatial averaging emerges from the approximation made for the direct correlation function that makes mapping to actual systems not unproblematic. Attempts to solve this problem have been presented in Refs. 25, 30, and 31.

### 2.2 The equation of motion

Considering that the particle density is a conserved field, an overdamped conservative dynamics is assumed in the PFC model, realized by the dimensionless equation

$$\frac{\partial \psi}{\partial t} = \nabla^2 \frac{\delta \Delta F}{\delta \psi} + \zeta. \quad (5)$$

Here $\delta \Delta F/\delta \psi$ is the functional derivative of the dimensionless free energy difference with respect to the reduced number density. The thermal fluctuations are represented by a colored Gaussian noise, $\zeta$, of a correlator $\langle \zeta(\mathbf{r}, t)\zeta(\mathbf{r}', t')\rangle = -\alpha \nabla^2 g(|\mathbf{r}-\mathbf{r}'|,\sigma)\delta(t-t')$, while $\alpha$ is the noise strength and $g(|\mathbf{r}-\mathbf{r}'|,\sigma)$ a high frequency cutoff function[23,26] for wavelengths shorter than the inter-particle spacing, $\sigma$. Eq. (5) follows from the equation of motion of the DDFT after making a few simplifications.[26,29] As a result of the assumed diffusive dynamics, the PFC models with this type of equation of motion are appropriate for crystal aggregation in colloidal systems.

A few remarks are appropriate here regarding the noise added to the equation of motion: In the classical DDFT-type models, nucleation does not occur in a homogeneous liquid unless Langevin noise, which represents the thermal fluctuations, is added to the equation of motion. While this procedure leads to nucleation, it is, however, not without conceptual difficulties, as discussed in the literature.[32-34] Considering the number density as an ensemble averaged quantity, all fluctuations are (in principle) incorporated into the free energy. Adding then noise to the equation of motion part of the fluctuations would be counted twice.[32,33] In contrast, if the number density is considered as a time averaged quantity, there is phenomenological motivation to incorporate noise into the equation of motion.[34] The latter standpoint is rather appealing practically: fluctuation (noise) driven crystal nucleation takes place indeed in the liquid, and capillary waves appear at the crystal-liquid interface. In recent PFC studies of crystal nucleation on the atomistic scale, a conserved noise term is used in the equation of motion [see Eq. (5)]. For this purpose, colored noise obtained by filtering out the unphysical short wavelengths (those that are smaller than the inter-particle distance) is often used.

### 2.3 The Euler-Lagrange equation

The extrema/saddle points of the (grand) free energy functional can be found by solving the respective Euler-Lagrange equation, which reads as

$$\frac{\delta \Delta \Omega}{\delta \psi} = \frac{\delta \Delta F}{\delta \psi} - \left.\frac{\delta \Delta F}{\delta \psi}\right|_{\psi_0} = 0. \quad (6)$$

Here $\Delta \Omega$ is the relative grand free energy, $\psi_0$ the reduced particle density of the reference liquid, $\delta \Delta F/\delta \psi|_{\psi_0} = \mu_0$ is the respective chemical potential, while periodic boundary condition is applied at the borders of the simulation box. In the case of the single-mode PFC model, the following form of the Euler-Lagrange equation applies:

$$[-\varepsilon + (1+\nabla^2)^2](\psi - \psi_0) = -(\psi^3 - \psi_0^3). \quad (7)$$

Eq. (6) together with the boundary condition represents a 4th order boundary value problem. The Euler-Lagrange equation has been used to determine the equilibrium properties of the single-mode PFC model, including the phase diagram, the solid-liquid interface free energy in 2D, the density difference at the solid-liquid interface in 2D, and the nucleation barrier for bcc and fcc structures in 3D.

### 2.4 Numerical methods

Owing to the higher-order differential operators one meets in the PFC models, solutions to the equation of motion and Euler-Lagrange equation are usually obtained numerically, relying on a pseudo-spectral successive approximation scheme combined with the operator-splitting method. A similar approach based on a spectral semi-implicit scheme[35] relying on parallel Fast Fourier Transform proved numerically highly efficient in solving the equation of motion, while assuming periodic boundary condition at the perimeter. GPU (Graphics Processing Unit) cards turned out to be highly efficient in solving the related problems.

### 2.5 Phase diagrams

As pointed out recently, the single- and two-mode PFC models realize DLVO-type potentials. The respective phase diagrams corresponding to 2D and 3D are shown in Fig. 3. In 2D, a single crystalline phase (the triangular phase) forms, which coexists with the homogeneous fluid and striped phases; a phase diagram similar to those predicted for weakly charged colloids.[36] In contrast, in 3D, additional stability domains occur for the bcc, hcp, and fcc structures, besides the 3D extensions (rods and the lamellae) of the respective 2D periodic structures. Remarkably, rod and lamellar structures, and a phase diagram akin to the phase diagram of the single-mode PFC have been observed in MD simulations performed using a DLVO-type potential.[37]

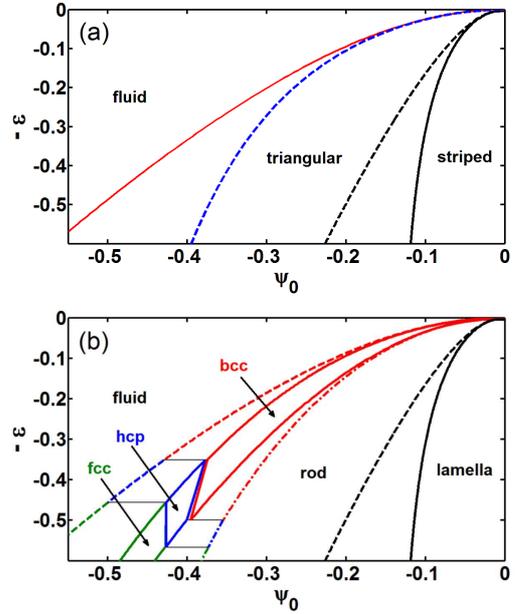

**Fig. 4** Phase diagrams of the single-mode PFC model used in addressing heterogeneous nucleation of nanocrystals in (a) 2D, and (b) 3D. Note the stability domains for the bcc, hcp, and fcc structures in the latter.

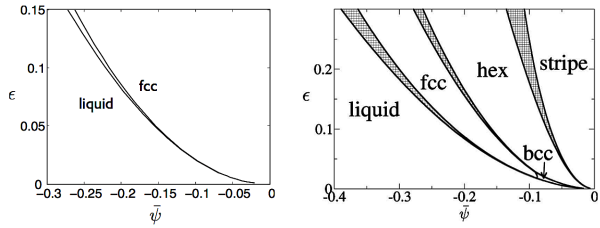

**Fig. 5** Phase diagrams of the two-mode PFC model used in addressing heterogeneous nucleation of nanocrystals: (a) Single-mode approximations to the phase diagram of the two-mode PFC model in 3D for $R_1 = 0$. (b) The same for $R_1 = 0.05$. Note the small bcc stability domain near the critical point. (Reprinted with permission from K.-A. Wu, A. Adland and A. Karma, *Phys. Rev. E*, 2010, **81**, art. no. 061601 © 2010 American Physical Society.)

The two-mode PFC model by Wu *et al.*[30], that has been designed to realize fcc crystallization, suppresses the bcc phase [Fig. 5(a)]. Interpolating between the full fcc ($R_1 = 0$) and the single-mode limits in terms of the parameter $R_1$ leads to the appearance of a bcc stability domain in the neighborhood of the critical point [Fig. 5(b)]. Whether the bcc stability domain is accompanied with an hcp stability domain, as seen in the single-mode limit, is yet unclear.

## 2.6 Homogeneous crystal nucleation

Before reviewing the results for heterogeneous crystal nucleation of nanocrystals, it is desirable to recall some essential findings concerning homogeneous crystal nucleation revealed by PFC investigations.

Having specified the free energy functional, nucleation (homogeneous or heterogeneous) can be addressed in two ways: (i) either via solving the Euler-Lagrange equation under the appropriate boundary conditions one determines the properties of the critical fluctuation (nucleus); or (ii) by solving the equation of motion with noise representing thermal fluctuations one simulates nucleation. Route (i) is fully consistent with the free energy functional. However, owing to the noise applied in the case of route (ii), the free energy of the phases change together with the phase diagram and the interfacial properties. Decreasing the noise amplitude, results from route (ii) should converge to route (i). As will be shown below, the full richness of the nucleation pathways can only be revealed by applying both routes.

*(a) Finding the properties of nuclei (solving the Euler-Lagrange equation):* The Euler-Lagrange equation method has recently been used to find the properties of heterophase fluctuations. Tóth *et al.*[38] has performed such study at a reduced temperature that leads to a faceted Wulff shape. The particle density of the liquid has been varied so that the size of the nuclei changed substantially. The initial guess for the solution of the Euler-Lagrange equation has been constructed so that a shape (e.g., cube sphere, octahedron, rhombo-dodecahedron) has been chosen, which was then filled with the analytic solution obtained using the single-mode analytic solution for the bulk crystal. This has been then placed on the background of the homogeneous liquid of particle density $\psi_0$, and a *tanh* smoothing has been performed at the perimeter. The Euler-Lagrange equation has been solved numerically with this initial guess. The size of the crystallite in the initial guess has been varied in small steps.

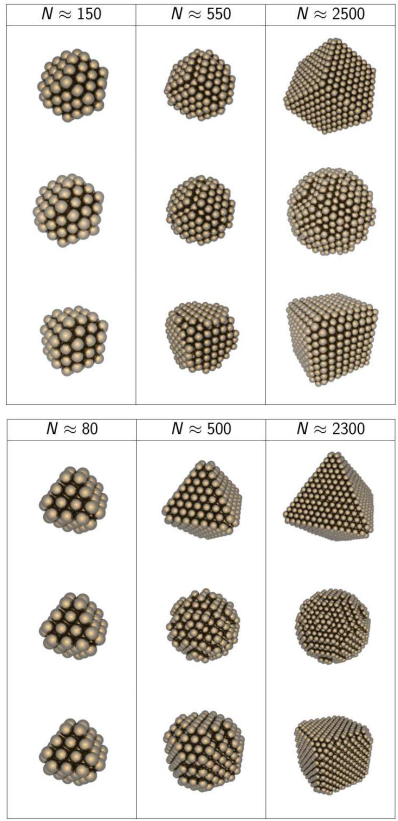

**Fig. 6** Equilibrium nanoclusters of (a) bcc and (b) fcc types found by solving the Euler-Lagrange equation starting from rhombic-dodecahedral/octahedral (top row), spherical (center), and cubic (bottom row) initial crystal shapes in the single-mode PFC model.

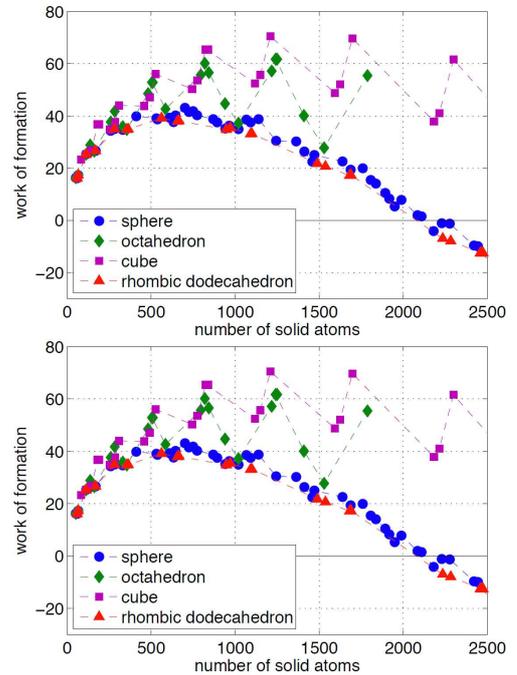

**Fig. 7** Dimensionless free energy of formation (made dimensionless as the free energy) for equilibrium nanoclusters of (a) bcc and (b) fcc structure as a function of size predicted the Euler-Lagrange equation of the single-mode PFC model.[38] Note the similar height of the nucleation barrier for the two structures. (Obtained at reduced temperature $\varepsilon = 0.3748$.) (Reprinted with permission from G. I. Tóth, G. Tegze, T. Pusztai, G. Tóth and L. Gránásy, *J. Phys. Condens. Matter*, 2010, **22**, art. no. 364101 © 2010 Institute of Physics.)

Contrary to the coarse-grained van der Walls/Cahn-Hilliard /Landau type models, where the nucleus is the only solution, here a very large number of cluster variants exist, that represent local minima of the free energy, which are all solutions of the Euler-Lagrange equation for $\psi_0$ in the far field (Fig. 6). This implies that the free energy surface is fairly rough. (A similar behavior has been reported for 2D.)

The results obtained the Euler-Lagrange method for bcc and fcc clusters are summarized in Figs. 7(a) and 7(b), respectively.[38] If the initial cluster shape is not compact, the higher free energy minima are found than for the compact shapes. In accordance with this, the spherical and the rhombic-dodecahedral (bcc) and octahedral (fcc) shapes provide the best guess for the minima in the free energy surface. The obtained free energy values indicate that the nucleation barrier is comparable for the bcc and fcc structures. This together with the similarity of the thermodynamic driving forces for fcc and bcc crystallization[27] at this specific value of $\varepsilon$ indicates that interface free energies for bcc and fcc structures are rather similar; a finding in a good agreement with direct computations for the interface free energies for flat interfaces.[39] On the other hand, these results appear to be in direct contradiction with those for metals from molecular dynamics simulations, which predict a significantly smaller interface free energy for the bcc phase.[40] It is worth noting, however, that the molecular dynamics results refer to low melting entropy materials, whose solid-liquid interface is rough / diffuse on the atomistic scale, as opposed to the high melting entropy corresponding to our system of strongly faceted sharp interface. Faceted interfaces are often associated with covalent type bonding, where a broken-bond model is usually a reasonable approximation. This approach yields comparable interface free energies for the bcc and fcc structures.[41] Apparently, the PFC results are consistent with earlier findings for faceted interfaces from the broken-bond model, while further work is warranted to clarify the disagreement between the PFC and MD predictions. We note that in obtaining these results it has been presumed that the crystalline phase nucleates directly from the supersaturated liquid. Time dependent simulations indicate, however, that this is often not the case, as complex nucleation pathways via a metastable precursors might turn out to be preferable.

*(b) Dynamic investigations (solving the equation of motion):* The 2D simulations by Gránásy et al. for the single-mode PFC ($\lambda = 1$) model indicate that at small supersaturations crystallization starts with direct nucleation of the triangular phase from the melt, whereas at large supersaturations formation of an amorphous precursor precedes crystal nucleation that takes place in the amorphous precursor.[42] The precursor has typical amorphous structural properties.

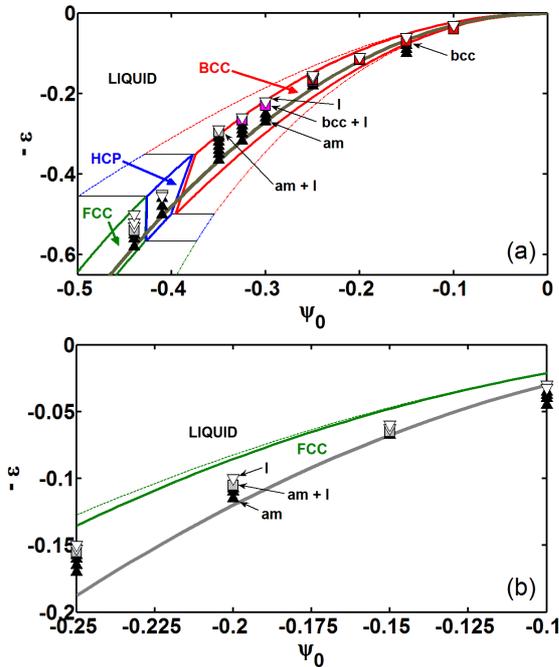

**Fig. 8** Homogeneous nucleation maps for PFC models characterized by DLVO-type pair potentials:[23] (a) single-mode PFC and (b) two-mode PFC. The state corresponding to $10^5$ time steps is shown: open triangle – liquid; square – amorphous + liquid; circle – amorphous + bcc; diamond – bcc; filled triangle – amorphous. The gray line indicates the linear stability limit of the liquid. The respective phase diagrams are also shown. (Reprinted with permission from G. I. Tóth, T. Pusztai, G. Tegze, G. Tóth and L. Gránásy, *Phys. Rev. Lett.*, 2011, **107**, art. no. 175702 © 2011 American Physical Society.)

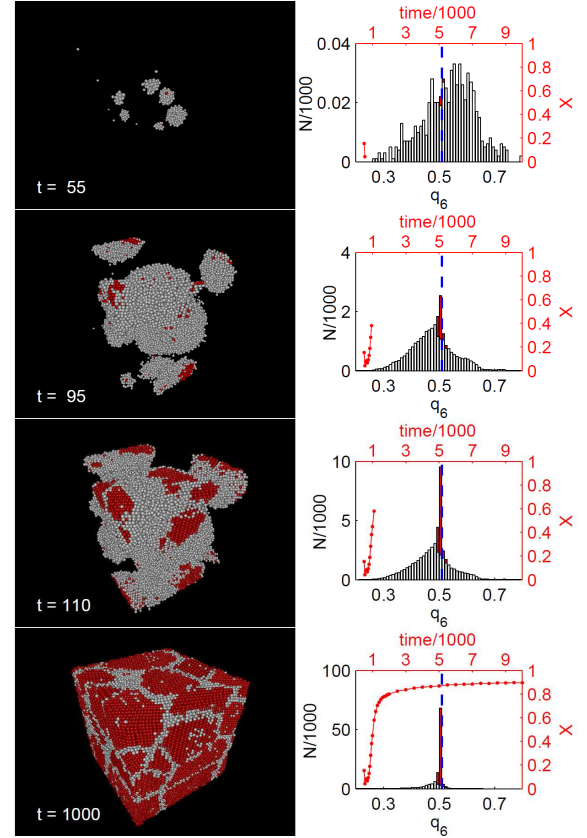

**Fig. 9** Two-step nucleation in the single-mode PFC model (characterized by DLVO-type pair potential[23]) at $\varepsilon = 0.1667$. Left: Snapshots of the particle density taken at dimensionless times are shown. Spheres of the diameter of the interparticle distance centered on density peaks higher than a threshold are shown that are colored red if $q_4 \in [0.02, 0.07]$ and $q_6 \in [0.48, 0.52]$ (bcc-like) and white otherwise. Right: Population distribution of $q_6$ (histogram painted similarly) and the time dependence of the fraction $X$ of bcc-like neighborhoods (dots and solid line). (Reprinted with permission from G. I. Tóth, T. Pusztai, G. Tegze, G. Tóth and L. Gránásy, *Phys. Rev. Lett.*, 2011, **107**, art. no. 175702 © 2011 American Physical Society.)

A similar behavior has been reported for 3D by Tóth et al.[23] Starting from a homogeneous fluid state, isothermal treatments ($\varepsilon$ = const.) have been performed for $10^5$ time steps. During the mapping of phase selection in the single-mode PFC model, several densities have been chosen within the bcc stability domain, and a single value in the hcp and the fcc domains. For each reduced density, a series of simulations has been conducted at several reduced temperatures, of which the lowest has been chosen so as to yield the amorphous phase, while the highest the liquid. Similar investigations have been performed for the two-mode PFC model in the fcc stability domain. The results of these investigations are summarized in Fig. 8. The only crystalline phase seen to nucleate in these dynamic studies is the bcc one, inside the bcc stability domain. There with decreasing reduced temperature the state after $10^5$ time steps varies as follows: liquid, bcc, bcc + amorphous, and amorphous. In other cases, coexistence of the amorphous and liquid phases is seen: liquid, amorphous + liquid, and liquid, corresponding to decreasing reduced temperatures.[23]

The kinetics of bcc nucleation has been further investigated within the framework of the single-mode PFC at the reduced particle density $\psi_0 = -0.25$ and reduced temperature $\varepsilon = 0.1667$ (cf., the melting point of $\varepsilon \approx 0.1475$.) The results are summarized in Fig. 9, where the left panels visualize the particle density field, while the right ones show the crystalline fraction as a function of time together with a histogram displaying the probability of neighborhoods characterized by the bond-order parameter $q_6$. The solid phase is composed of pronounced density peaks, whereas only small-amplitude fluctuations are present in the liquid. Only the peaks that exceed a threshold are displayed, for which spheres of the atomic radius have been drawn. The spheres have been colored according to the local values of the $q_4$ and $q_6$ rotationally invariant order parameters that monitor the local order around a particle. [For definition of the bond order parameters see Appendix A and Ref. 43. In the case of perfect crystals $q_6$ = 0.575 (fcc); 0.485 (hcp); 0.511 (bcc) and 0.354 (sc).] If $q_4 \in$ [0.02, 0.07] and $q_6 \in$ [0.48, 0.52] (bcc-like) the sphere is painted red, while the rest of the "atoms" is colored white. The histogram in the right panels shows the population of neighborhoods characterized by $q_6$. The fraction of particles of bcc-like neighborhood (the red ones) is also displayed, representing the bcc fraction $X$ of the particles (see the red curve in the right panels).

In this study, first amorphous clusters form (nucleate) and grow, on which the bcc structure nucleates subsequently (bcc nucleation on an amorphous precipitate). The simulation box fully solidifies into a polycrystalline state with amorphous grain boundaries [Fig. 8(d)]. These findings strongly indicate that crystal nucleation is enhanced by the presence of the amorphous precursor, and that bcc crystal nucleation directly from the liquid phase requires several orders of magnitude longer time than via the precursor. This behavior appears analogous to the non-crystalline precursor assisted crystal nucleation in colloidal systems[20] and simple liquids.[21,22] Remarkably, the amorphous phase coexists with the liquid, and nucleates from the liquid phase (separate amorphous "drops" form).[23] These suggest that the amorphous precursor is a mestastable phase that forms from the liquid by a first-order phase transition as indeed reported previously by Berry et al.[44] Further support the for amorphous precursor mediated mechanism is given by the "average" bond order parameter maps, $\bar{q}_4$ - $\bar{q}_6$ and $\bar{q}_4$ - $\bar{q}_6$.

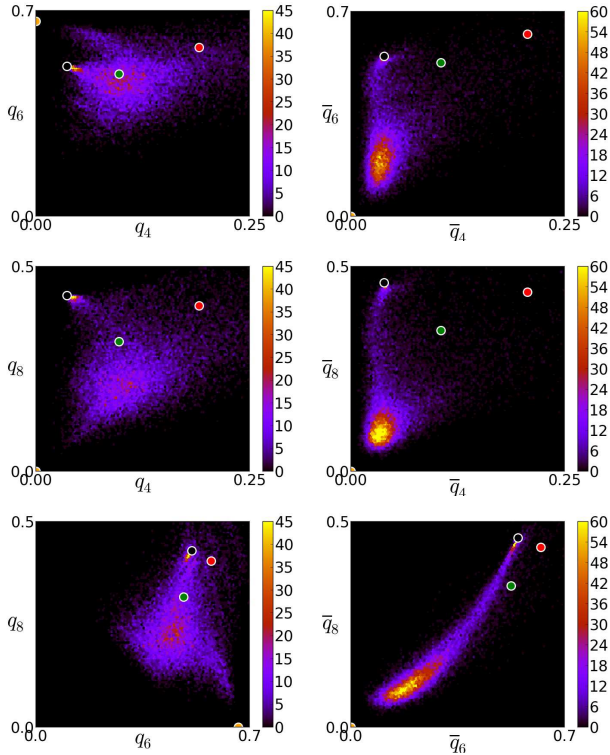

**Fig. 10** Bond order parameter maps and average bond order parameter maps for an intermediate stage ($t$ = 90) of the two-step nucleation shown in Fig. 8. (a), (b) $q_4 - q_6$ and $\bar{q}_4$ - $\bar{q}_6$; (c), (d) $q_4 - q_8$ and $\bar{q}_4$ - $\bar{q}_8$; while (e), (f) $q_6 - q_8$ and $\bar{q}_6$ - $\bar{q}_8$. The circles stand for ideal structures: black – bcc; green – hcp; red – fcc; and yellow – icosahedral. Comparison with molecular dynamics simulations for the Lennard-Jones system[43] indicates that the amorphous precursor formed in the single-mode PFC has structural properties similar to the simple liquids. Note that the structure of the liquid cannot be analyzed in the PFC model as there the particle density is essentially constant with a small amplitude noise.

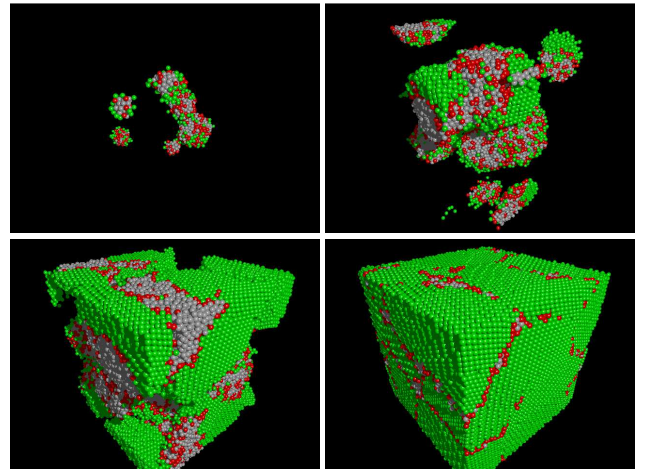

**Fig. 11** Modified Kawasaki-Tanaka type coloring of the particles in the PFC simulation shown in Fig. 8: grey if $\bar{q}_6$ < 0.28, red if $\bar{q}_6 \in$ [0.28, 0.4], green if $\bar{q}_6$ > 0.4. It appears that (i) owing to the time averaging inherent in the PFC model, this model cannot detect medium range crystalline order, and (ii) the bcc phase forms on the surface of the amorphous regions, much like heterogeneous nucleation. Time elapses from left to right and from top to bottom.

(These average bond order parameters by Lechner and Dellago give a larger separation between different structures.[43]) One observes initially the formation of the amorphous structure accompanied with a subsequent appearance of the bcc structure at intermediate times (see Fig. 10). The specific coloring (red if $\bar{q}_6 \in [0.28, 0.4)$, and green if $\bar{q}_6 \in [0.4, 0.55]$) Kawasaki and Tanaka[19] have used to visualize medium range crystalline order indicates, as expected, that due to the time averaging inherent in density functional type approaches, medium range crystalline order cannot be observed in the PFC model (Fig. 11). Red-colored particles appear only at the interface between the well localized particles in the crystal (green) and in the amorphous phase $\bar{q}_6 < 0.28$, which we paint grey. The snapshots indicate furthermore that the bcc phase appears on the surface of the amorphous phase, much like in the case of heterogeneous nucleation.

Even in the case of the two-mode PFC model, which is specifically designed to promote crystallization to the fcc phase, no trace of fcc nucleation has been observed.[23] A detailed analysis in terms of the respective driving forces (grand potential density difference with respect to the liquid) and the fcc-liquid and glass-liquid interface free energies shows that structure evolution and density change are decoupled, and in the temperature/density range accessible for dynamic simulations, the nucleation of the density change (amorphous freezing) is faster than the nucleation of the fcc phase (structural change). This follows from the finding that the free energy of the glass-liquid interface is about 2/3 of the fcc-liquid interface. It is, nevertheless, clear from the thermodynamic data that (analogously to the 2D case) at small undercoolings/supersaturations, there is a regime, where direct crystal nucleation from the liquid should take place; however, there the time for nucleation is prohibitively long for dynamic simulations.

In a recent analysis the preference for bcc nucleation in these PFC models has been attributed to a specific form of the effective pair potential evaluated from the structural data for the amorphous phase: for both the single- and two-mode PFC models, besides a minimum at $r_0$, the pair potential has a maximum at $\sim r_0\sqrt{2}$, and weaker minima further outside.[23] Such potentials are known (i) to suppress fcc and hcp crystallization,[45] and has been identified as a possible source for the lack of hcp and fcc nucleation in dynamic simulations performed using the equation of motion,[23] whereas (ii) the multiple minima are expected to lead to coexisting disordered phases.[46]

Summarizing, the PFC models display metastable amorphous-liquid coexistence and first-order liquid to amorphous transition.[23] In the domains, where crystallization is accessible for dynamic simulations, the nucleation of the amorphous phase is faster than crystal nucleation. This leads to a separation of time scales for density and structural changes, as seen in several other systems (hard sphere and Lennard-Jones systems, and 2D and 3D colloids). However, some details might differ: The amorphous-liquid coexistence is unknown in the hard sphere system, while the fcc and hcp structures are suppressed in the PFC models. It is also unclear whether along the reaction coordinate specified in Ref. 21, the free energy landscape of the PFC models is indeed similar to that of the Lennard-Jones system. Combining the results obtained for various potentials, it appears that a repulsive core suffices for the appearance of a disordered precursor, whereas the peak at $\sim r_0\sqrt{2}$ correlates with the observed suppression of fcc and hcp structures, while the coexistence of the liquid and amorphous phases seen here can be associated with multiple minima of the interaction potential.[23] Remarkably, similar amorphous-precursor mediated bcc nucleation has been reported for an extended PFC model with parameters fitted to Fe (see also Electronic Supplementary Information).[38,42]

## 3. Heterogeneous nucleation of nanocrystals in the single-mode PFC model

Several aspects of heterogeneous crystal nucleation require atomistic studies. Herein, results of PFC modeling on the structural aspects of the substrate-crystal interaction will be reviewed, such as the effects of lattice mismatch and the structure of the substrate on the nucleation barrier, the contact angle, and surface adsorption of the crystalline phase. In presenting the findings, we follow the route used in the case of homogeneous nucleation: First, the results obtained by the Euler-Lagrange equation are to be addressed, followed by results from dynamic studies based on solving the equation of motion. As in the case of homogeneous nucleation, the two methods prove complementary. In the works summarized below, the *crystalline substrate* is represented by a periodic potential term $V(\mathbf{r})\psi$, added to the free energy density [to the integrand of Eq. (2)].[24] Here $V(\mathbf{r}) = [V_{s,0} - V_{s,1}S(a_s, \mathbf{r})]h(\mathbf{r})$, where $V_{s,0}$ controls crystal adsorption, $V_{s,1}$ is the amplitude of the periodic part, $S(a_s, \mathbf{r})$ is a single-mode solution function that provides the periodic structure of the substrate,[26] $a_s$ the lattice constant of the substrate, whereas $h(\mathbf{r}) \in [0, 1]$ is an envelope function defining the size and shape of the substrate.[24] Note, furthermore, that the anisotropy of the crystal-liquid interface decreases towards the critical point for both 2D and 3D.[26]

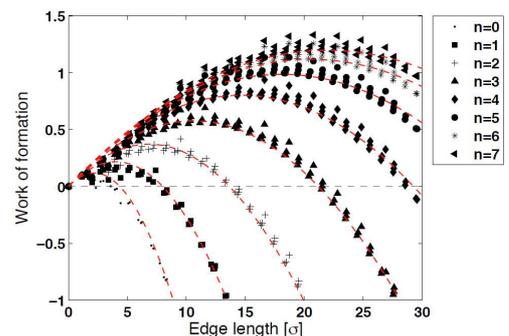

**Fig. 12** Dimensionless nucleation barrier for heterogeneous nucleation (made dimensionless as the free energy) vs. size relationship obtained by solving the Euler-Lagrange equation for faceted nuclei in 2D. The supersaturation decreases with increasing $n$. The lattice constant of the substrate is equal to the interparticle distance in the triangular crystal. The lines are to guide the eye. Here, "edge length" is the length of the free side of the crystallite parallel with the substrate (see Fig. 13(e)). (Reprinted with permission from G. I. Tóth, G. Tegze, T. Pusztai, G. Tóth and L. Gránásy, *J. Phys. Condens. Matter*, 2010, **22**, art. no. 364101 © 2010 Institute of Physics.)

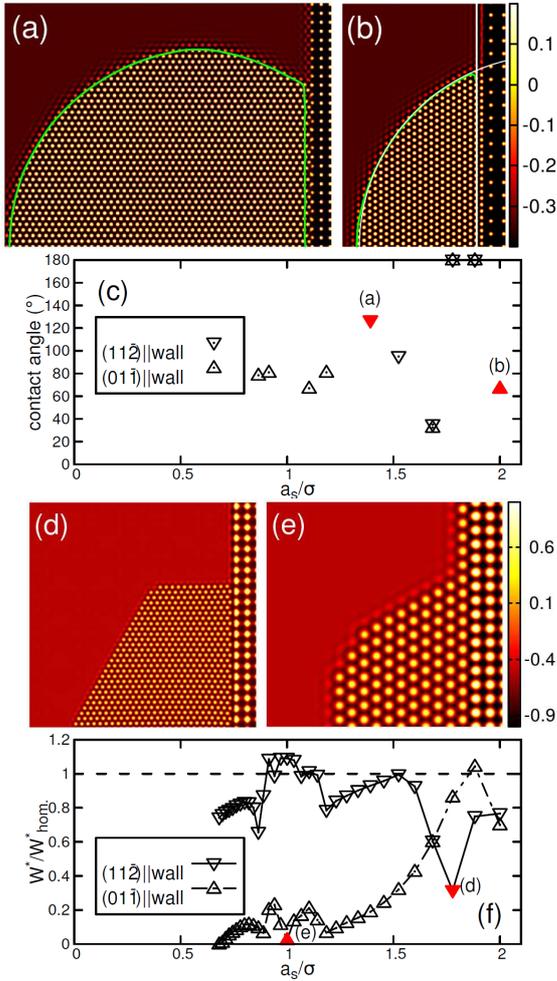

**Fig. 13** Heterogeneous nucleation on a flat wall in 2D from solving the Euler-Lagrange equation for a square lattice substrate.[24]

(a),(b) Typical (nonfaceted) nuclei obtained for small anisotropy. Here $a_s/\sigma = 1.49$ and 2.0, respectively, while the orientations are (11-2) and (01-1) parallel with the wall. The intersection of the circular and linear fits (white lines) to the contour line (green) defines the contact angle. (c) Contact angle versus $a_s/\sigma$ for small anisotropy. The full triangles stand for cases shown in panels (a) and (b).

(d),(e) Faceted nuclei obtained far from the critical point, at $a_s/\sigma = \sqrt{3}$ and 1.0. Respective orientations: (11-2) and (01-1) parallel with the wall. (f) Work of formation of faceted nuclei normalized by the value for homogeneous nucleation ($W^*=W^*_{hom}$) vs $a_s/\sigma$. The full triangles stand for cases shown in panels (d) and (f).

(Reprinted with permission from G. I. Tóth, G. Tegze, T. Pusztai and L. Gránásy, *Phys. Rev. Lett.*, 2012, **108**, art. no. 025502 © 2012 American Physical Society.)

### 3.1 Results by the Euler-Lagrange equation

*(a) Nucleation on flat surfaces:* First, results for heterogeneous nuclei forming in 2D on a flat square-lattice wall of varied lattice constant are presented. Here, the free energy surface has many local minima allowing the Euler-Lagrange equation to map out the nucleation barrier (see Fig. 12).[38] Two dominant relative orientations have been observed in dynamic simulations: faces (01-1) or (11-2) parallel with the wall.

*Determination of contact angle:* The misfit dependence of the contact angle has been first evaluated at a relatively *weak anisotropy*.[24] Here $\theta$ is defined as the angle between the linear and circular parts of the closed contour line corresponding to $(\psi_L + \psi_S)/2$ in the coarse-grained (filtered) particle density [see Figs. 13(a) and 13(b)]. (Subscripts S and L denote the solid and liquid phases.) A nonmonotonic relationship between the contact angle and the reduced lattice constant $a_s/\sigma$ of the substrate has been reported [Fig. 13(c)]. Here $\sigma$ is the interparticle distance in the 2D triagonal structure.

In the case of *strong anisotropy* yielding faceted interfaces far from the critical point, the contact angle is apparently determined by the crystal structure and orientation: The contact angle $\theta$ is 60º when the orientation (01-1) is parallel to the wall [Fig. 13(d)], whereas it is 90º when the orientation (11-2) is parallel to the wall [Fig. 13(e)], independently of the monolayer occasionally seen to form on the wall.

*Nucleation barrier:* As for the homogeneous case,[38] the work of formation of the equilibrium clusters fits well to the classical $W(l) = Al^2 + Bl$ relationship, where $l$ is the linear size of the nucleus (Fig. 12).[38] Accordingly, the nucleation barrier ($W^*$) has been defined as the maximum of the fitted formula. $W^*$ data obtained so for the two orientations are shown for $1/2 < a_s/\sigma \le 2$ in Fig. 13(f). Remarkably, the $W^*$ vs. $a_s/\sigma$ relationships are nonmonotonic, and have deep minima for the matching lattice constants ($a_s/\sigma = 1$ and $\sqrt{3}$ for the two orientations seen in dynamic simulations performed using the equation of motion). Except for extreme lattice mismatch, nuclei having the orientation (01-1) parallel with the wall dominate.

*(b) Nucleation on nanoparticles:* Next, we review the 2D results predicted for the *free-growth limited mechanism* of particle induced crystallization on a square-shaped nanoparticle, under relatively weak or fairly large aniso-tropies. To ensure nearly perfect wetting (a precondition of the free-growth limited model), $a_s = \sigma$ has been set. (This is not ideal wetting as the crystal structures differ on the two sides of the wall-solid interface.) Two linear sizes have been chosen for the study: $L_s = 4\sigma$ and $L_s = 32\sigma$.

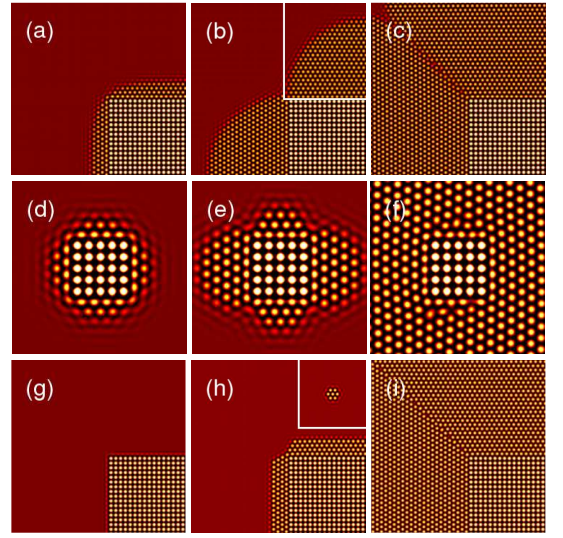

**Fig. 14** Free-growth-limited mode of particle induced crystallization on square shaped square-lattice substrates as predicted by the Euler-Lagrange equation in 2D.[24] The liquid density, the reduced temperature, and the size have been changed as follows: (a)–(c) $\varepsilon = 0.25$ and $L_s = 32\sigma$. (d)–(f) $\varepsilon = 0.25$, and $L_s = 4\sigma$. (g)–(i) $\varepsilon = 0.5$ and $L_s = 32\sigma$. In all cases $a_s/\sigma = 1$. The supersaturation increases from left to right. The inserts show the corresponding homogeneous nuclei. Note that (i) in all cases there is a critical supersaturation beyond which free growth takes place, and that (ii) small clusters are more faceted under the same conditions than the large ones (*cf.* (a) & (b) and (d) & (e)). (Reprinted with permission from G. I. Tóth, G. Tegze, T. Pusztai and L. Gránásy, *Phys. Rev. Lett.*, 2012, **108**, art. no. 025502 © 2012 American Physical Society.)

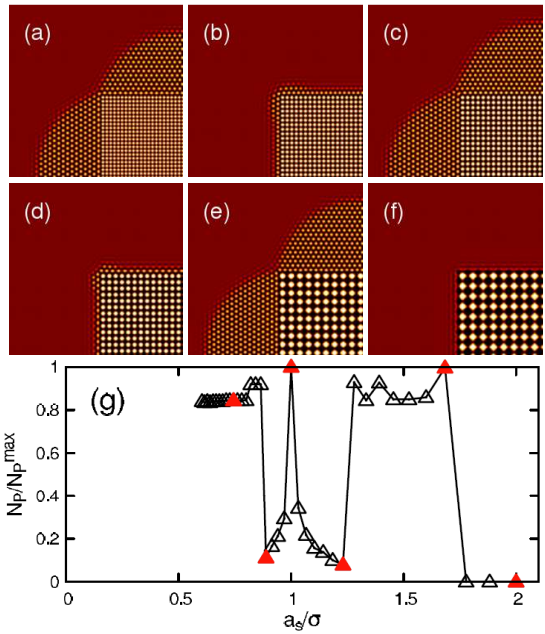

**Fig. 15** Adsorption of the crystalline phase on square-shaped particles versus mismatch at small anisotropy.[24] (a)–(f) These are equilibrium states obtained by solving the Euler-Lagrange equation for $a_s/\sigma$ increasing from left to right and from top to bottom. (g) Number of adsorbed crystalline particles normalized by their maximum vs the reduced lattice constant. The full triangles stand for results corresponding to panels (a)–(f). (Reprinted with permission from G. I. Tóth, G. Tegze, T. Pusztai and L. Gránásy, *Phys. Rev. Lett.*, 2012, **108**, art. no. 025502 © 2012 American Physical Society.)

The results for the *larger nanosubstrate* ($L_s = 32\sigma$) obtained assuming a relatively weak anisotropy indicate that even outside of the coexistence region adsorbed crystal layers form on the surface of the substrate [Fig. 14(a)], which evolve into circular "caps" inside the coexistence region [Fig. 14(b)]. When the diameter of the homogeneous nucleus becomes smaller than $L_s$, free growth commences [Fig. 14(c)].[24] This observation is in excellent agreement with the free-growth limited model. For the *smaller size*, however, a faceted crystal shape is observed, and the free-growth limit is reached at a monatomic critical size that is much smaller than $L_s$ [Fig. 14(d)-(f)]. At large distance from the critical point, faceted crystals form [Fig. 14(g)-(i)]. Here, free growth takes place, when the critical size is much smaller than $L_s = 32\sigma$. These findings indicate that the free-growth limited mechanism is valid so far as the foreign particles are sufficiently large, and the free energy of the solid-liquid interface has only a weak anisotropy.

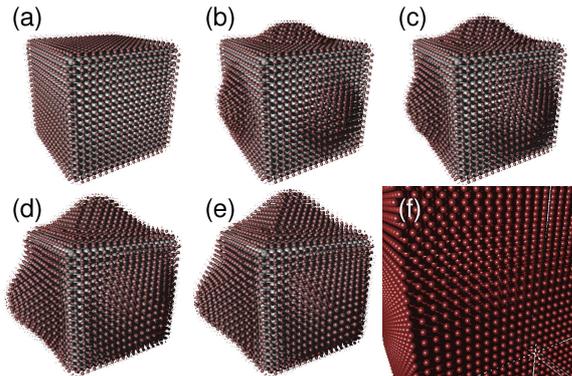

**Fig. 16** Free-growth-limited mode of particle induced crystallization in 3D on a cube shaped particle of simple cubic structure. Here $\varepsilon = 0.25$ and the supersaturation changes from left to right and from top to bottom, $L_s = 16 a_{bcc}$, where $a_{bcc}$ is the lattice constant of the stable bcc structure. (The Euler-Lagrange equation has been solved on a $256 \times 256 \times 256$ grid.) Spheres centered on the number density peaks are shown, whose size increases with the height of the peak. Color varies with the height of the density peak, interpolating between red (minimum height) and white (maximum height). (Reprinted with permission from G. I. Tóth, G. Tegze, T. Pusztai and L. Gránásy, *Phys. Rev. Lett.*, 2012, **108**, art. no. 025502 © 2012 American Physical Society.)

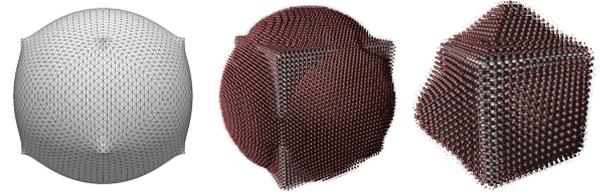

**Fig. 17** Stable shape preceding free growth in the free-growth-limited mode of particle induced crystallization on square-shaped square-lattice substrates. (a) Theoretical shape for infinite size,[11(b)] and PFC predictions for (b) $L_s = 64\sigma$ and (c) $L_s = 32\sigma$. Note that with decreasing size a faceted shape develops as reported for homogeneous nucleation by Backofen and Voigt.[47] (Leftmost panel reprinted with permission from S. A. Reavley and A. L. Greer, *Philos. Mag.*, 2008, **88**, 561–579. © 2008 Taylor & Francis.)

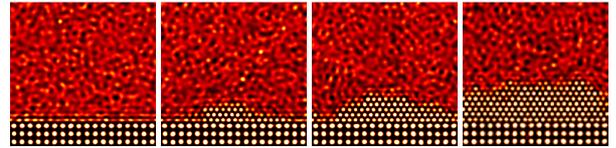

**Fig. 18** Heterogeneous nucleation on a flat substrate of square-lattice structure in the single-mode PFC model (obtained by solving the equation of motion). Time elapses from left to right.

Next the *effect of lattice mismatch on the adsorption* of the crystalline phase is presented for 2D.[24] The lattice constant of the substrate has been varied between $\sigma/2$ and $2\sigma$, so that it stays commensurable with $L_s = 32\sigma$. The results are summarized in Fig. 15. The amount of crystalline phase adsorbed on the particle is a nonmonotonic function of $a_s$. At $a_s = \sigma$ nearly semi-circular crystal adsorbates appear on the faces of the nanocrystal substrate [see Fig. 15(c)], whereas for slightly different $a_s$ much thinner crystal layers are observed on both sides [Fig. 15(g)]. Further away from the nearly perfect fit ($a_s = \sigma$), the adsorbed layer thickens again; yet for very large mismatch (such as $a_s \approx 2\sigma$), crystal adsorption is forbidden.

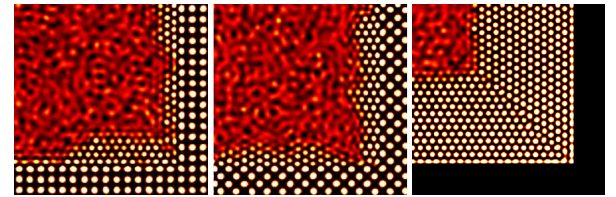

**Fig. 19** Heterogeneous nucleation in 2d in rectangular inner corners as predicted by the single-mode PFC model.[43] (a) Nucleation on (01) surfaces of a square lattice (ratio of lattice constant of substrate to interparticle distance $a_s/\sigma \approx 1.39$). (b) Nucleation on (11) surfaces of a square lattice. (c) Nucleation on an unstructured substrate. Note the frustration at the corner and the formation of a grain boundary starting from the corner at later stages.

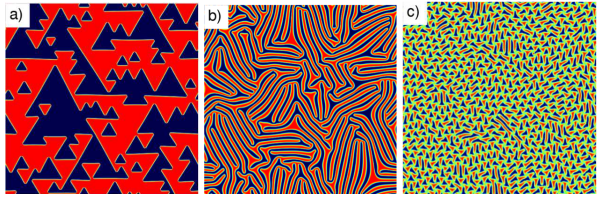

**Fig. 20** Surface patterns predicted by a PFC model for Cu monolayer on Ru (0001) surface. The coupling between the layer and the substrate decreases from left to right. Coloring: fcc domains are blue, hcp domains are red, and the domain walls are green. (Reprinted with permission from K. R. Elder, G. Rossi, P. Kanerva, F. Sanches, S. C.Ying, E. Granato, C. V. Achim and T. Ala-Nissila, *Phys. Rev. Lett.*, 2012, **108**, art. no. 226102. © 2012 American Physical Society.)

Testing of the *free-growth limited model* has been extended to 3D,[24] using a cube shaped foreign particle of simple cubic (sc) structure and of $a_s$ that coincides with the interatomic distance of the bcc structure. The investigations have been performed in the stability domain of the bcc structure. The results are in a *qualitative* agreement with the free-growth limited model (Fig. 16).

It is remarkable, however, that the morphology of the adsorbed crystalline layer preceding free growth depends strongly on the size of the substrate. The shape for the case, where the interface thickness is negligible relative to the linear size of the substrate has been obtained by Reavley and Greer [Fig. 17(a)],[11(b)] whereas the PFC solutions at the critical supersaturation obtained for $L_s = 64\sigma$ and $L_s = 32\sigma$ are shown in Figs. 17(b) and 17(c). While the first is a continuously curving surface, in the case of the $L_s = 64\sigma$ cubic cluster spherical caps form on (100) faces, whereas in the case of the $L_s = 64\sigma$ cluster pyramids form on the same faces.

### 3.2 Results from the equation of motion

*(a) Nucleation on flat surfaces & corners:* Gránásy *et al.*[42] investigated 2D crystal nucleation on flat walls and in rectangular corners of structured and unstructured substrates within the single-mode PFC. In the case of a flat square-lattice substrate a relatively small reduced undercooling implying a small anisotropy and sufficient mismatch to prevent immediate growth from the surface of the substrate ($a_s/\sigma = 1.39$) have been employed. A sequence of snapshots showing the formation of clusters via 2D heterogeneous nucleation and the late stage growth morphology are displayed in Fig. 18. Remarkable are the large amplitude capillary waves and the continuous appearance/disappearance of pre-nucleation clusters during the initial stage of crystallization.

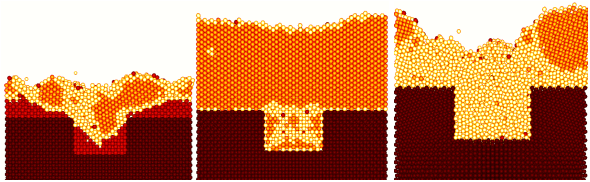

**Fig. 21** Crystallization on fcc substrate with a rectangular nanoscale pit (equation of motion in 3D).[24] Spheres drawn around density peaks larger than a threshold are shown. Order parameters $q_4$ and $q_6$ have been used for the structural analysis. Hues changing from dark to light stand for the substrate, and the fcc, bcc, and amorphous structures, respectively. ($\varepsilon = 0.16$ and $\psi_0 = -0.25$) From left to right $a_s/a_{fcc} = 1.0$, 1.098, and 1.42. Cross-sectional views are displayed. (Reprinted with permission from G. I. Tóth, G. Tegze, T. Pusztai and L. Gránásy, *Phys. Rev. Lett.*, 2012, **108**, art. no. 025502 © 2012 American Physical Society.)

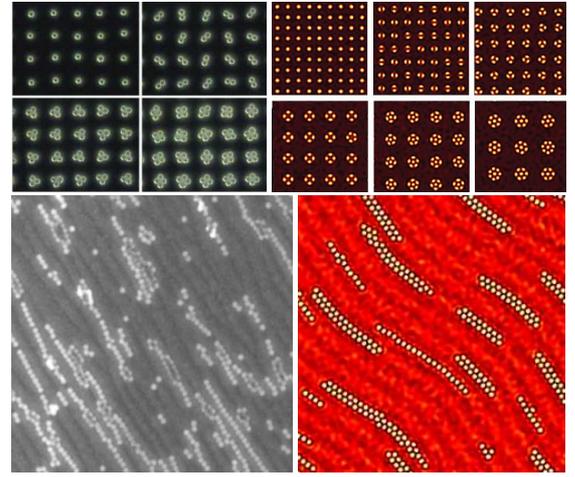

**Fig. 22** Colloid patterning in experiment[50] (left) and dynamic PFC simulations (right). Upper row: occupation of chemically patterned substrate (experimental image reprinted with permission from I. Lee, H. Zheng, M.F. Rubner and P. T. Hammond, *Adv. Mater.*, 2002, **14**, 572–577. © 2002 WILEY-VCH Verlag GmbH, Weinheim, Fed. Rep. of Germany). Lower row: pattern formation due to immersion capillary forces on a rippled substrate surface (experimental image reprinted with permission from A. Mathur, A. Brown, and J. Erlebacher, *Langmuir*, 2006, **22**, 582–589. © 2006 American Chemical Society).

2D crystal nucleation in rectangular corners of structured and unstructured substrates has also been investigated.[42] In spite of expectation based on the classical theory or conventional PF simulations,[6(a)] in which the corners are preferred nucleation sites, in the PFC model the rectangular corner does not appear to assist crystal nucleation owing to the misfit of the triangular crystal structure with a rectangular corner (Fig. 19). Crystals of different orientation nucleate on the two substrate surfaces, which leads to the formation of a grain boundary starting from the corner. The free energy of forming the grain boundary makes the rectangular corner a non-favorable place of nucleation. A 60° corner, in turn, favors the nucleation of the triangular phase.

*(b) Nucleation on nano-patterned surface:* A possible way to influence crystallization is to use the crystal lattice of the substrate to influence pattern formation on its surface, a problem addressed recently.[48,49] A binary extension of the PFC model, supplemented by a periodic external field, has been used to map the effect of coupling strength on pattern formation at the surface (Fig. 20).[49]

The effect of lattice mismatch has been investigated for crystallization initiated by an fcc substrate with rectangular pit (see Fig. 21):[24] For matching $a_s$ values, fcc and bcc epitaxy have been reported, however, with interference with edge-induced frustration. At high lattice mismatch, amorphous-phase-mediated bcc crystallization occurs, an analogue of the two-step mode of homogeneous nucleation.

### 3.3 Future directions

The PFC studies could be further extended to explore the effect of various nanoscale features of the substrate on nucleation, including surface roughness, surface curvature, chemical patterning of the surface, etc. Interesting effects can be addressed once the PFC method is combined with fluid flow. Steps have been made recently in this direction. Recent advances in PFC modeling offers ways to model real materials.[25,26,30,31]

Apparently, strategies established in colloid patterning might be profitably employed for governing the crystallization process on the nanoscale (via e.g., chemical patterning or via nanocapillary forces).[26] It might be then expected that the PFC methodology, that has been worked out for modeling colloid patterning[42] (see Fig. 22), can also be employed on the nanoscale.

## 4. Summary and concluding remarks

In this Tutorial Review, we have given a brief overview of recent developments in Phase-Field Crystal modeling of heterogeneous nucleation of/on nanocrystals. The results compiled here extend previous knowledge in several directions:

(i) At large supersaturations, homogeneous nucleation of the stable crystalline phase is expected to happen via an *amorphous precursor*; a kinetically preferred pathway emerging from a time scale separation of the density and structural changes.

(ii) The *lattice mismatch* between the substrate and the crystal influences nonmonotonically such properties as the contact angle, the thickness of the crystalline layer adsorbed on the substrate, and the height of the thermodynamic barrier for heterogeneous nucleation.

(iii) The highly successful *free-growth limited model* of particle-induced crystallization by Greer and coworkers[11] is valid for larger nanoparticles ($L_s = 32\sigma$) and small anisotropy of the solid-liquid interface free energy, whereas for small nanoparticles ($L_s = 4\sigma$) or faceted crystals, the critical supersaturation, beyond which free growth takes place, substantially deviates from the one predicted by analytic theory.

(iv) A large mismatch between the crystal and the substrate may lead to an amorphous surface layer, which assists the formation of the crystalline phase; a heterogeneous analogue of the amorphous precursor mediated homogeneous crystal mode.

### Acknowledgements

This work has been supported by the EU FP7 Project "EXOMET" (contract no. NMP-LA-2012-280421, co-funded by ESA), and by the ESA MAP/PECS project "MAGNEPHAS III".

## Appendix: Structural Characterization: Bond Orientational Order

In order to characterize the local structure around a particle $k$, Steinhardt et al.[43(a)] have introduced the rotationally invariant bond order parameters

$$q_l^k = \left\{ \frac{4\pi}{2l+1} \sum_{m=-l}^{l} \left| q_{lm}^k \right|^2 \right\}^{1/2},$$

where

$$q_{lm}^k = 1/n_b^k \sum_{j=1}^{n_b^k} Y_{lm}(\mathbf{r}_{kj}).$$

Here $Y_{lm}(\mathbf{r}_{kj})$ are the spherical harmonic functions of degree $l$, and order $m$, and $n_b^k$ is the number of the bonds of particle $k$. Recently, Lechner and Dellago[43(b)] have introduced a coarse-grained version extended to the second neighbors:

$$\bar{q}_l^k = \left\{ \frac{4\pi}{2l+1} \sum_{m=-l}^{l} \left| \bar{q}_{lm}^k \right|^2 \right\}^{1/2},$$

and

$$\bar{q}_{lm}^k = 1/N_b^k \sum_{j=0}^{N_b^k} q_{lm}^j.$$

In the latter expression, the sum for $j$ runs for all neighbors $N_b^k$ of particle $k$ including the particle itself. Accordingly, in computing the average $\bar{q}_l^m$ for particle $k$, one uses the local orientational order vectors averaged over particle $k$ and its surroundings. Note that $q_l^k$ relies on structural information from the first shell around particle $k$, whereas in its averaged version $\bar{q}_l^k$ structural information from the second shell is also taken into account. This spatial averaging has a tremendous significance in detecting local ordering with high sensitivity: In Fig. 10, we compare the original and the coarse-grained bond order parameter maps. Indeed, the separation of the structures is far more pronounced in terms of the average bond order parameters.